\documentclass[10pt,twocolumn]{article}

\usepackage[letterpaper,margin=0.75in]{geometry}
\usepackage{times}
\usepackage{graphicx}
\usepackage{amsmath}
\usepackage{amssymb}
\usepackage{authblk}
\usepackage{caption}
\usepackage{cite}
\usepackage{hyperref}
\usepackage{soul}
\usepackage{xcolor}
\usepackage{array}

\title{\vspace{-1em}
Reply to Smallenburg: Near-melting nucleation and the exponential growth of hard-sphere nucleation times
}

\author[1]{Roseanna N. Zia\thanks{Corresponding author: \texttt{your.email@missouri.edu}}}
\affil[1]{Department of Mechanical and Aerospace Engineering, University of Missouri, Columbia, MO 65211, USA}
\affil[ ]{{\small correspondence: rzia@missouri.edu}}

\date{}

\begin{document}

\twocolumn[
\maketitle

\begin{center}
\begin{minipage}{0.88\textwidth}
Smallenburg reports {\em near-melting point} simulations and observes a well-predicted spontaneous nucleation with mixed, finite-time morphology, in a Comment on our recent Perspective~\cite{wang2026elusive}. The Comment's incorrect broader takeaway --- that equilibrium coexistence is ``readily achievable'' --- rests on an untested generalization from a single state point near the phase envelope, and misses entirely the intriguing role played by Frenkel's underlying mechanism. We reiterate the salient point missed by the Comment: the nucleation time grows astronomically with just tenths of a percent of volume fraction away from 53\%.  This phenomenology emerges from the entropy exchange mechanism Frenkel described, which predicts that spontaneous phase separation is dynamically accessible only down to about 53\% volume fraction from the melting point, and astronomically long waiting times through most of the remaining phase envelope. We provide here calculations to address the potential misconception created by the Comment.
\end{minipage}
\end{center}

\vspace{1em}
]

The absence of reports of spontaneous, non-induced coexistence in pristine simulations of monodisperse, purely-repulsive hard spheres (MPRHS) {\em except nearly at the melting point} was presented in our recent Perspective~\cite{wang2026elusive} as the expected consequence of Frenkel's entropy-exchange mechanism\cite{frenkel1993order} and the associated nucleation barrier. At high density near melting, spontaneous nucleation can become observable on accessible simulation times, as our companion paper demonstrated earlier this year~\cite{wang2026frenkel}, and Smallenburg's Comment reinforces. As one moves away from that favorable regime, the driving force for nucleation decreases, the barrier rises sharply, and the waiting time increases by many orders of magnitude. Indeed, the Comment's own simulations reinforce the phenomenology and the question at hand. Smallenburg's reported simulations are performed at a single packing fraction, $\phi=0.5325$, deliberately chosen in the narrow high-density portion of the coexistence region where spontaneous nucleation is most expected to become computationally feasible  under Frenkel's nucleation-barrier framework~\cite{frenkel1993order}. However, the Comment's assumption of equivalent behavior farther from the melting point is incorrect and, unfortunately, misleading. We clear this up here.

It is straightforward to translate Frenkel's Lennard-Jones atomic system to a colloidal system to predict the behavior at, and close to, $\phi=53\%$ and nearby, as follows. Frenkel’s estimate of the nucleation barrier predicts astronomically long waiting times until near $\approx20\%$ supersaturation, at which the typical experimental atomic nucleation rate is 1 nucleus per $\mathrm{cm}^3$ per second.  {This means that to establish a spontaneous, stable phase-separated state, a simulation of a $10^{-16}\mathrm{cm}^3$ box containing $10^6$ particles would need to continue for approximately  {$10^{16}$s} (about $3.17\times10^8$ years) of laboratory time (which, even with today's far faster compute capabilities, will be longer simulation time) ~\cite{tenwolde1996numerical}}. The supersaturation in ten Wolde {\em et al.}’s system varies with temperature and, correspondingly, the ratio of that distance from the melting point relative to the distance of the melting point from absolute zero, $(T_m-T)/(T_m-T_0 )$. Here, $T_0=0 K$ where motion ceases. In colloidal systems, the analogous supersaturation is the distance from the hexagonal closed packed condition (HCP), where Brownian motion is assumed to vanish, as:
\begin{equation*}
\textrm{\%supercooling}=\frac{(\phi-\phi_F)}{(\phi_{HCP}-\phi_F )}.
\end{equation*}
Using $\phi_F=0.4918$, $\phi_{HCP}=0.74$, and $\phi_M=0.543$, the 20\% supercooling condition corresponds to $\phi=0.541$. Independent calculations cited by ten Wolde et al. yield comparable values of $\phi=0.536$ \cite{kelton1991crystal}. 

 {Nucleation is faster in a larger system and faster for colloids, but the same exponential growth is predicted in the $\sim$ 20\% supersaturation region.} Given the volumetrically larger system size used in our {\em colloidal} scale simulation ($1.6 \times 10^{-8} \mathrm{cm}^3$ with 2,048,000 particles of size 200nm), we estimate that a laboratory system of the same size would require two years of real time to produce one nucleus at these volume fractions (the time required in simulation will as always, be longer). This estimate follows  {the reference nucleation rate of} 1 nucleus per $\mathrm{cm}^3$ per second  {using Frenkel's atomic conditions described the previous paragraph.}  {While the two-year laboratory time estimate with {\em atomic} nucleation rates is still prohibitively long for simulations}  {near 53\%,}  {our results showed spontaneous phase separation and long-term coexistence above that, at $\phi=0.535$ \cite{wang2026frenkel}, which Smallenburg also reproduced and reported in his Comment at a nearby single point $\phi=0.5325$.}  {This illustrates two points. First, nucleation rate is incredibly sensitive in colloids near melting, growing dramatically with just a few tenths of a percentage into the supersaturation region. Second, nucleation in a colloidal system is faster than in the hard-sphere atomic system -- orders of magnitude, although still growing exponentially, which Frenkel subsequently pointed out: In a Monte Carlo study of pristine MPRHS, Auer and Frenkel\cite{auer2004numerical} show}  {the reduced nucleation rate $I^*=I\sigma^5/D_0$ at $\phi=0.5342$ (17.1\% supercooling) is $10^{-9}$, where $\sigma$ is the size of particle and $D_0$ is the particle's self diffusion coefficient. For typical parameters at colloidal scale, the size $\sigma$ is $200$nm, dynamic viscosity of water $\eta$ is $0.001\mathrm{Pa}\cdot\mathrm{s}$, and temperature $T$ is $300$K. This leads to the dimensional nucleation rate}  {at $\phi=0.5342\%$} of  {$I=6.9\times10^{6}$ nuclei per $\mathrm{cm}^3$ per second.} Given our simulation box size ($1.6 \times 10^{-8} \mathrm{cm}^3$ with 2,048,000 particles),  {the physical nucleation time needed to be covered is calculated to be 9.1sec in the lab, or a few hours in simulations. This expected, more accessible laboratory timescale can be completed in hours to days in simulation, {\em matching our results}\cite{wang2026frenkel} as well as the ones reproduced in Smallenburg's Comment.}

Now, Frenkel's study further shows that the nucleation time in MPRHS colloids increases by about 6-fold per 0.1\% volume fraction decrease~\cite{auer2004numerical}. For example,  {using the baseline calculation at $\phi=0.5342$~\cite{auer2004numerical},} the nucleation time for colloidal simulations at $\phi=0.53$ is three laboratory hours  {(which translates to years in simulations, even on today's high-performance supercomputers)} but balloons to 16,000 years at $\phi=0.52$. 

The Table in \textbf{Figure \ref{fig:timetable}} shows the laboratory time for nucleation and that it grows exponentially marching lower in volume fraction from $\phi\approx0.53$, predicting that spontaneous, non-induced phase separation remains extraordinarily slow throughout most of the coexistence region for MPRHS. Only above $\approx0.53$ does the high propensity for crystallization make spontaneous phase separation dynamically accessible without bias or triggers. Thus, Smallenburg's Comment addresses the obvious and favorable limiting case rather than the broader phenomenological question emphasized in the Perspective~\cite{wang2026elusive}.

\begin{table}[t!]
\centering
\begin{tabular}{| >{\centering\arraybackslash}m{0.08\linewidth} | >{\centering\arraybackslash}m{0.24\linewidth} | >{\centering\arraybackslash}m{0.24\linewidth} | >{\centering\arraybackslash}m{0.24\linewidth} |} 
 \hline
 $\phi$ & Nucleation time in sim (sec) & Nucleation time in sim ($a^2/D_0$) & Nucleation time in sim (sec, hrs, yrs) \\ 
 \hline\hline
 0.540 & 0.0002 & 0.044 & near inst \\ [0.1em]
 \hline
 0.535 & 1.5 & 300 & 1.5 sec \\[0.1em]
 \hline
 0.530 & $1.0\times10^4$ & $2.1\times10^6$ & 2.9 hrs \\[0.1em]
 \hline
 0.525 & $7.2\times10^7$ & $1.5\times10^{10}$ & 2.3 years \\[0.1em]
 \hline
 0.520 & $5.0\times10^{11}$ & $1.0\times10^{14}$ & 16,000 years \\ [0.1em]
 \hline
 0.515 & $3.5\times10^{15}$ & $7.0\times10^{17}$ & 110M years \\ [0.1em]
 \hline
 0.510 & $2.4\times10^{19}$ & $4.8\times10^{21}$ & 765B years \\ [0.1em]
 \hline
 0.505 & $1.7\times10^{23}$ & $3.3\times10^{25}$ & never \\ [0.1em]
 \hline
 0.500 & $1.2\times10^{27}$ & $2.3\times10^{29}$ & never \\ [0.1em]
 \hline
 0.495 & $8.0\times10^{30}$ & $1.6\times10^{33}$ & never \\ [0.1em]
 \hline
\end{tabular}
\caption{Table showing exponential growth in nucleation time in a system of 2,000,000 purely repulsive, hard-sphere colloids {, calculated based on a linear fit to the data from Figure 11 in Ref.~\cite{auer2004numerical}}. }
\label{fig:timetable}
\end{table}

Overall, the atomic and colloidal theory predicts nucleation and spontaneous phase separation growing exponentially slow once volume fraction moves just a few tenths of a percent in volume fraction below $0.53$, growing from a few seconds for colloids (a few years for atomic systems) to thousands, then millions, then billions of years. This sensitivity near $\phi\approx 0.53$ is reflected in Smallenburg's spotty results for the final mixture, which are reminiscent of Alder and Wainwright’s early results: space-spanning crystallites formed under finite-size constraints~\cite{alder1957phase, alder1959studies, alder1960studies}. A closer look at Smallenburg's data reveals that this distinction is reinforced by  his own sampling: only 8 out of 50 slab simulations and 11 out of 50 cubic simulations produced any crystallization at all, and the scattered values of final crystal fraction, from 56\% to 100\%,  do not establish that the 80\% theoretical value{, which is calculated based on lever rule,} has been achieved consistently or spontaneously. {It is possible the variation in final crystal fraction is due to finite size effects (despite nearly $10^5$ particles in the Smallenburg simulations) or the finite simulation duration.  Examination of Smallenburg's Fig. 1(a) suggests that for these simulations to converge to the theoretical final crystal fraction of 80\%, much longer simulation durations will be required}.    {The Comment  therefore reinforces} what the Perspective article \cite{wang2026elusive} addressed and our original results\cite{wang2026frenkel} predicted: $\phi \approx 0.53$ is the point of departure to exponentially slowing nucleation.  {We acknowledge that [1] was unclear that the extremely long time scales for nucleation apply to {\em most} of the coexistence regime (not at melting), but the time scales become dramatically and usefully smaller very close approach to the melting point. As shown here, very near 20\% supersaturation (in colloids, this is very near the melting point), phase separation becomes finitely attainable in both atomic and colloidal simulations and, in both systems, phase separation time increases exponentially moving away from about 53\% volume fraction, becoming practicably unattainable without biasing techniques just a few tenths of a percent of volume fraction away.}

We hope the reader can appreciate the consequential phenomenology explained by Frenkel: {moving away from the melting point further into the coexistence region leads to an astronomical increase in the nucleation barrier as the freezing line is approached}, and hence to Frenkel's prediction of $3.17\times10^8$ years for full phase separation in atomic systems~\cite{tenwolde1996numerical} not too far from 20\% supersaturation and even larger farther away. As noted above, this exponential increases is also predicted in colloids by Auer and Frenkel (2004)\cite{auer2004numerical}.

One of our original points was that, for various reasons, no-one had reported a simulation in which spontaneous emergence of two coexisting phases emerged. We tested Frenkel's hypothesis, showed the sensitivity of the departure point near $\phi=0.53$, and predicted phase separation there. We are delighted that since publication of the Perspective, Smallenburg has been able to reproduce our finding, and we are pleased that our work nucleated interest in this topic and efforts to fill this interesting gap in the literature with understanding and further results.

This productive interchange highlights the complementary perspectives brought by molecular phase behavior in the physics community and colloidal phase behavior in the engineering community. Importantly, the authors of the Comment and of this Response have discussed the broader merit of the work in showing how these kinetic challenges are not merely obstacles but instead reveal the underlying structure of Frenkel's entropy-exchange mechanism~\cite{frenkel1993order}, specifically, how the delicate balance between configurational and vibrational entropy becomes dynamically inaccessible under pristine conditions. We view this as an opportunity to bring renewed attention to an underappreciated aspect of hard-sphere phase behavior. Overall, we see this dialogue as a useful step toward aligning perspectives and refining how coexistence and kinetics are framed in the literature, and we welcome the opportunity to continue that exchange here.

\bibliographystyle{unsrt} 

\end{document}